%% file: main.tex
\documentclass[sigconf,natbib=true]{acmart}
\usepackage{xspace}
\usepackage{subcaption}
\usepackage{caption}
\usepackage{multirow}
\usepackage{color}

\newcommand{\ie}{\emph{i.e.,}\xspace}
\newcommand{\eg}{\emph{e.g.,}\xspace}

\newcommand{\paratitle}[1]{\smallskip\noindent \textbf{#1}\xspace}
\newcommand{\modelname}{DIL\xspace}

\AtBeginDocument{
  }

\setcopyright{acmcopyright}
\copyrightyear{2018}
\acmYear{2018}
\acmDOI{XXXXXXX.XXXXXXX}

\acmConference[Conference acronym 'XX]{Make sure to enter the correct
  conference title from your rights confirmation emai}{June 03--05,
  2018}{Woodstock, NY}
\acmPrice{15.00}
\acmISBN{978-1-4503-XXXX-X/18/06}

\begin{document}

\title{Retraining A Graph-based Recommender with Interests Disentanglement}

\author{Yitong Ji}
\affiliation{%
  \institution{Nanyang Technological University}
  \country{Singapore}
}
\email{s190004@e.ntu.edu.sg}

\author{Aixin Sun}
\affiliation{%
  \institution{Nanyang Technological University}
  \country{Singapore}
}
\email{axsun@ntu.edu.sg}

\author{Jie Zhang}
\affiliation{%
  \institution{Nanyang Technological University}
  \country{Singapore}
}
\email{zhangj@ntu.edu.sg}

\begin{abstract}
In a practical recommender system, new interactions are continuously observed. Some interactions are expected, because they largely follow users' long-term preferences. Some other interactions are indications of recent trends in user preference changes or marketing positions of new items. Accordingly, the recommender needs to be periodically retrained or updated to capture the new trends, and yet not to forget the long-term preferences. In this paper, we propose a novel and generic retraining framework called Disentangled Incremental Learning (DIL) for graph-based recommenders. We assume that long-term preferences are well captured in the existing model, in the form of model parameters learned from past interactions. New preferences can be learned from the user-item bipartite graph constructed using the newly observed interactions. In  DIL, we design an \textit{Information Extraction Module} to extract historical preferences from the existing model. Then we blend the historical and new preferences in the form of node embeddings in the new graph, through a \textit{Disentanglement Module}. The essence of the disentanglement module is to decorrelate the historical and new preferences so that both can be well captured, via carefully designed losses. Through experiments on three benchmark datasets, we show the effectiveness of DIL in capturing dynamics of user-item interactions. We also demonstrate the robustness of DIL by attaching it to two base models - LightGCN and NGCF.
\end{abstract}

\ccsdesc[500]{Information systems~Recommender systems}

\keywords{recommender system, incremental learning, streaming data}
\maketitle

\input{Intro.tex}

\input{Related.tex}

\input{gcnmodel.tex}

\input{dilModel.tex}

\input{Experiment.tex}

\input{discussion.tex}

\section{Conclusion}
\label{sec:conclude}
In this paper, we propose a novel and generic incremental learning strategy DIL for GCN-based recommender. The key idea of DIL is the pre-learning fusion. We extract long-term historical information from the model learned in previous period. Then historical information is fused with learnable new information. The embedding obtained after fusion is treated as initial feature of a node in a GCN-based recommendation model. To ensure that, the learnable new information could be trained to reflect new trends in the recommender system, disentanglement is used to ensure new information are independent to the historical information. By instantiating DIL on two GCN models, \ie LightGCN and NGCF, we show that DIL is effective in retraining and learn both short-term and long-term information for recommendation. The idea of ``pre-learning fusion'' provides an alternative way to consider how to extract and fuse historical information for model retraining. 
\bibliographystyle{ACM-Reference-Format}
\bibliography{reference}

\end{document}

%% file: Intro.tex
\section{Introduction}
\label{sec:intro}

\begin{figure*}[t]
    \centering
	  \begin{subfigure}[t]{0.6\columnwidth}
    	\centering
    	\includegraphics[width=\columnwidth, clip=True]{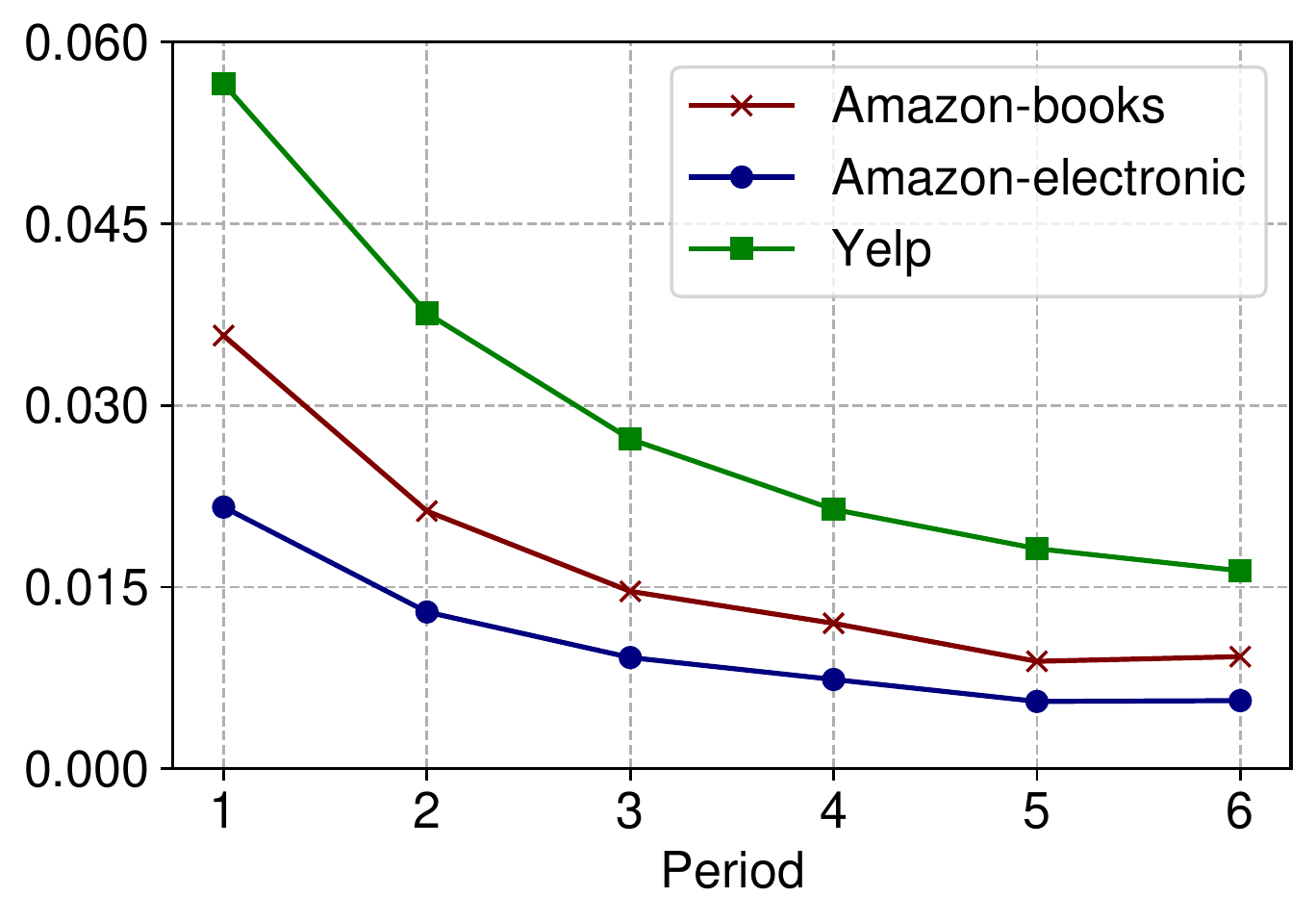}
    	\caption{LightGCN}
        \label{sfig:lightgcn_decay_performance}
	\Description{}
    \end{subfigure}
    \quad
	  \begin{subfigure}[t]{0.6\columnwidth}
    	\centering
    	\includegraphics[ width=\columnwidth, clip=True]{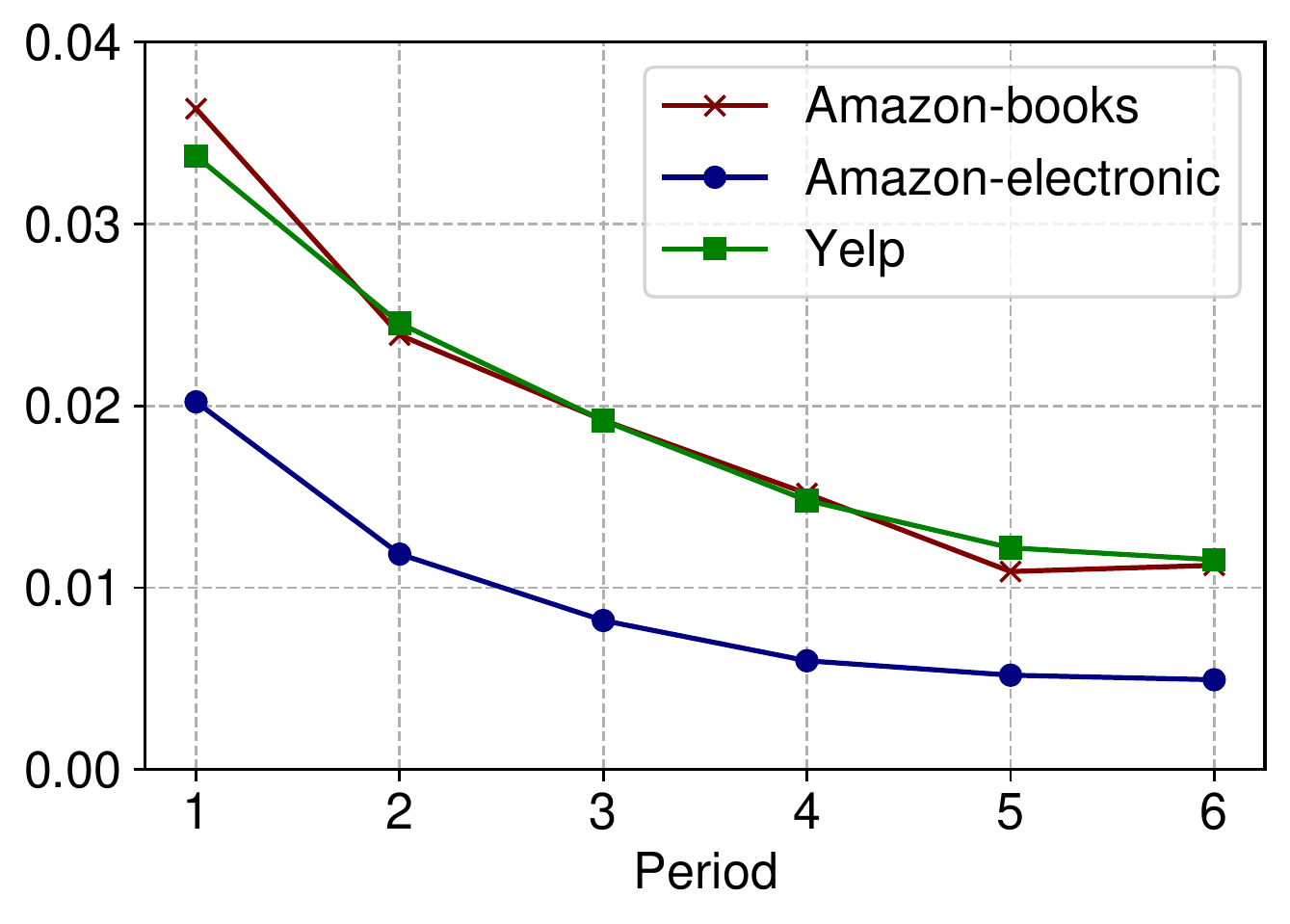}
    	\caption{NGCF}
        \label{sfig:ngcf_decay_performance}
	\Description{}
    \end{subfigure}
    \caption{Recall@20 over 6 periods in Amazon-books, Amazon-electronic and Yelp if there is no retraining conducted for LightGCN and NGCF.}
    \label{fig:decay_performance}
\end{figure*}

In the real-life context, a recommender system for a growing business is consistently expanding with increasing number of users and items. Moreover, interactions between users and items are continuously observed. To better serve its users, the recommendation model shall be updated from time to time, to keep track of the most recent updates, \eg new users, new items, and new trends indicated in recent interactions. The necessity of refreshing a recommendation model using the most recent interactions can be confirmed using Figure~\ref{fig:decay_performance}. We train LightGCN and NGCF using a specific amount of data from Amazon-books, Amazon-electronic and Yelp. \footnote{We conduct experiment on Amazon-books, Amazon-electronic and Yelp using data from Jan 2017 to June 2017, 2013 to 2014, and 2012 to 2014 respectively.} After training, LightGCN and NGCF is fixed without any form of retraining. Then, these fixed models are used for making recommendations over the next 6 periods, where a period refers to a different time interval depending on the dataset: one month for Amazon-books and half a year for Amazon-electronic and Yelp. Figure~\ref{fig:decay_performance} shows that the recommendation performance deteriorates over time. This decline is mainly due to the fact that an outdated model without retraining cannot effectively capture the latest trends in the recommender system. Therefore, it stresses the importance of updating the recommendation model when new interactions are observed. In a real-life platform equipped with recommender systems, retraining of the recommendation models are also conducted regularly~\cite{WideDeep16, Monolith22Byte}.

\begin{figure}[t]
    \centering
    \includegraphics[trim = {5.8cm 3.8cm 5.15cm 6.5cm}, clip, width = \columnwidth]{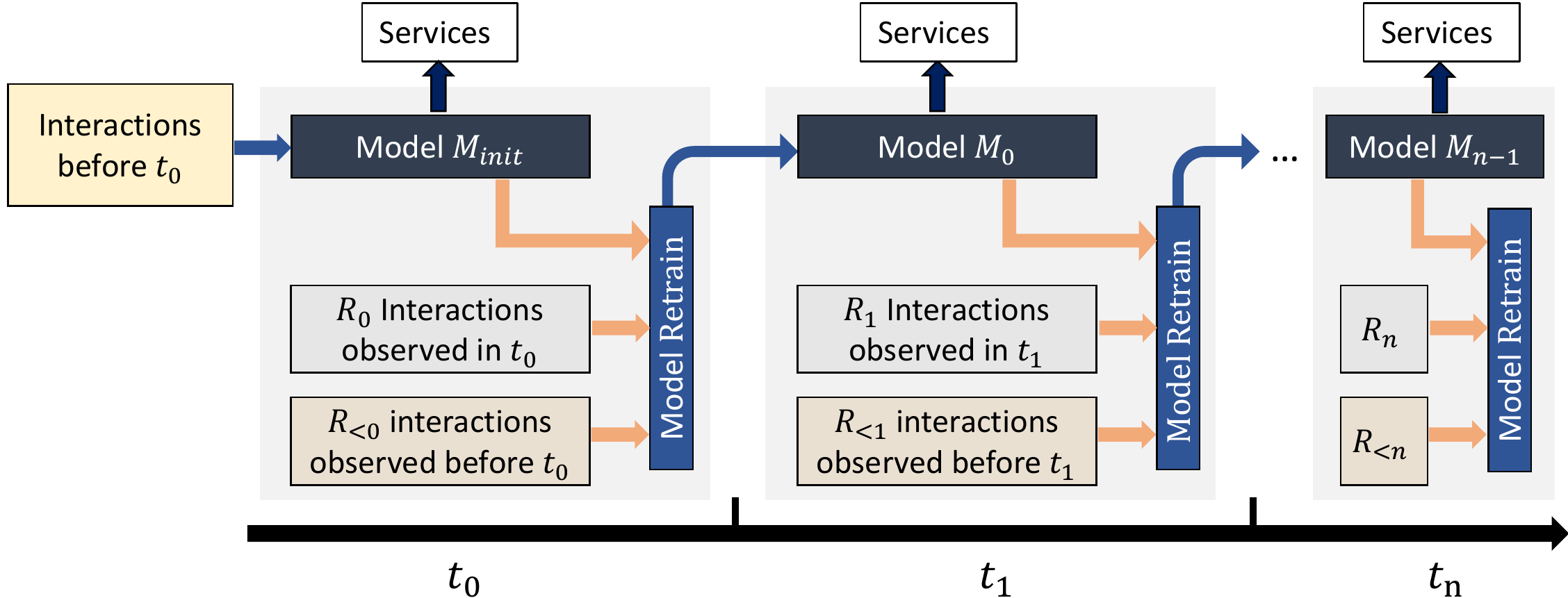}
    \caption{Incremental Learning with Model Retraining. The retraining may take inputs from (i) the current model, (ii) the newly observed interactions within the current time window, and (iii) historical interactions before this time window.}
    \label{fig:retraining_process}
\end{figure}

Without loss of generality, we consider the scenario that model retraining is conducted at some pre-scheduled time points. These time points can be scheduled periodically with a fixed frequency \eg a few months, or based on the number of newly observed interactions since the last model update.  Figure~\ref{fig:retraining_process} shows a typical retraining process, where  model retraining happens at the end of some time windows $t_0$, $t_1$, till $t_n$. To start with, there is an initial model $M_{init}$ learned from the interactions observed before $t_0$. Then $M_{init}$ is deployed for service during time window $t_0$. The model is retrained at the end of $t_0$, and the updated model $M_0$ will be deployed to service in $t_1$. At the end of any time window, \eg $t_n$, model retraining could utilize \textit{\textbf{three types of information}}: $R_n$, $R_{<n}$, and $M_{n-1}$. Here, $R_n$ is the set of new interactions observed during $t_n$. $R_{<n}$ refers to all historical interactions that are observed before $t_n$. 
$M_{n-1}$ is the latest model, that was updated in the previous time window $t_{n-1}$. Its parameters represent most updated knowledge captured before $t_n$ by the recommender, after the multiple updates since $M_{init}$.

The simplest retraining strategy is \textbf{fine-tune}. Using the latest model $M_{n-1}$ as the base model, fine-tune retrains $M_{n-1}$ with the newly observed interactions $R_n$, to get $M_{n}$.
However, this strategy comes with a potential problem - \textit{catastrophic forgetting}~\cite{overcomingCatastrophicForgetting, structureAwareExperienceReplay}. That is, the long-term preference of a user may be forgotten, because the model is forced to always focus on the current trends in each update. In Figure~\ref{fig:decay_performance}, it can be observed that although recall@20 decreases over time when there is no retraining of a recommendation model, it does not drop to $0$ even after 6 periods of no retraining. The fact that recall@20 has a non-zero value suggests that the user preferences and item characteristics may remain relatively stable over time, allowing an old model trained on historical interactions to accurately predict new interactions. Hence, historical information expressed in historical interactions should be preserved for more accurate recommendation. To fully utilize both historical and new information, the ``extreme'' strategy is \textbf{full-retrain}. Full-retrain abandons the current model and rebuilds the recommender from scratch, by using all interactions seen \ie $R_{<n} \cup R_n$. 
This solution has scalability issue because the training cost increases along time, with increasing number of interactions. More importantly, not all historical interactions carry relevant information to a recent context~\cite{loyalUser, optimalTrainingWindow, dynamicPopularity, graphsail}.

In light of the discussions above, existing studies~\cite{SML, ASMG, SPMF, graphsail, CI} resort to a scalable and effective retraining strategy - \textbf{Incremental Learning}. All incremental learning strategies fully utilize the newly observed interactions in the current time window. At the same time, they adopt different techniques to force the model to ``memorize'' old information, by using sampled old interactions, current model parameters, or both. For example, a subset of historical interactions are sampled in~\cite{SPMF, SSRM, GAG} as representative examples for long-term information. SML~\cite{SML} and ASMG~\cite{ASMG} treat model parameters, \eg user embeddings, learned in the past as historical information. 

In this paper, we propose an incremental learning strategy named \textbf{Disentangled Incremental Learning (\modelname)},  for Graph Convolutional Network (GCN) based recommender. We focus on GCN-based recommender for its capability in modeling complex relations between users and items, and its proven effectiveness~\cite{ngcf, lightgcn, neighborEnrichGraph, graphAugment}. 

Similar to any other incremental learning strategies, the two main challenges in designing \modelname are: (i) \textit{how to extract historical information}, and (ii) \textit{how to blend historical and new information} in the retrained model. Regarding the first challenge, some existing methods treat the entire node representations learned in the previous model as historical information~\cite{CI, IGCN}. In designing \modelname, we take a different approach. It is not necessary to fully take the previous node representations because not all information encoded in these representations are meant for long-lasting attributes. Further, as GCN network gathers information from different hops of neighbors to construct a node's representation, we treat information from neighbors at different hops separately. We consider that the neighbor nodes at different hops of the previous graph contribute differently to the historical information. Hence, we design an \textit{Information Extraction Module (IEM)} in DIL to dynamically adjust what and how much historical information to extract from the previous model. The key idea of IEM is to learn the weights in adjusting what and how much historical information to extract, rather than directly extracting some predefined piece of information from the previous model.  To address the second challenge in blending new and old information, we propose a \textit{Disentanglement Module (DM)}  to decorrelate the two types of information. DM ensures that the old and new information are treated differently in model retraining, to reflect their corresponding semantics. Yet, the two pieces of information work together through GCN learning to achieve accurate recommendations. 

The contributions made in this work are threefold. First, the proposed IEM represents \textit{a new approach of extracting historical information} from previous model, and we offer two implementation designs (Section~\ref{ssec:iem}). Second, we introduce \textit{disentanglement of historical information and new information} to ensure that both information could be well captured (Section~\ref{ssec:dm}). To the best of our knowledge, we are the first to introduce disentanglement of long-term preference and short-term preference in retraining of a recommendation model. Third, to combine the two modules, we are also the first to \textit{redesign the embedding layer} in GCN-based recommender for model retraining. An embedding represents a node's feature. It now contains both historical information by IEM and new information to be learned from the current interaction graph. Through message propagation in GCN, information exchanges among adjacent nodes and among indirect neighbors, leading to more accurate recommendation (Section~\ref{ssec:optim}). 
We instantiate DIL on two base models: NGCF~\cite{ngcf} and LightGCN~\cite{lightgcn}, and conduct experiments on three benchmark datasets: Amazon books, Amazon electronic, and Yelp. Through experiments, we show that DIL is effective in retraining GCN models. Moreover, it is generic for GCN-based model retraining and works well when augmenting to GCN-based recommendation models.

%% file: Related.tex
\section{Related Work}
\label{sec:related}
In this section, we first briefly review GCN-based models and then distinguish our \modelname from the existing retraining strategies. 

\subsection{GCN-based Recommendation}
\label{ssec:related.gcn}

Non-graph based recommendation models generally learn representations using observed user-item interactions. Other relations like user-user, item-item cannot be well modeled. NGCF~\cite{ngcf} is among the early models that exploit graph structure to capture these additional relations. It follows a GCN model and designs neighbor aggregation layer to draw information recursively from multi-hop neighbors. The model was then refined to a simpler version LightGCN~\cite{lightgcn}
after removing operations that do not contribute much to recommendation task. Drawing information from neighbors can be computationally costly, PinSage~\cite{pinsage} adopts random walk and graph convolutions for efficient representation learning in a large-scale recommender system. Despite the effectiveness of considering additional user-user, item-item relations, different neighbors could contribute different amounts of information for recommendation. KGAT~\cite{KGAT} employs attention mechanism in the neighbor aggregation layers to place different weights on neighbors. Unlike KGAT which distinguishes neighbors in the same hop, NIA-GCN~\cite{neighborsInteractionGCN} proposes to differentiate neighbors across different layers. Other than using a single user-item interaction graph, multi-GCCF~\cite{mgccf} further constructs a user-user graph and an item-item graph by using pairwise cosine similarity. These graphs are used as additional information for representation learning. The authors in~\cite{GNNSocialRec} propose to utilize user-user social graph as additional information for recommendation. To further enhance the robustness of a GCN-based recommendation model, SGL~\cite{SGL} adopts self-supervised learning in graph learning. Three graph augmentation techniques are proposed to generate multiple views of a graph for self-supervised learning. However, authors in~\cite{graphAugment} demonstrate that graph augmentation is less necessary. They propose to generate contrastive views by adding uniform noise to the node embeddings.

\subsection{Incremental Learning in Recommendation}
\label{ssec:related.incremental}

It is a common practice in the industry to retrain recommender system to capture the most recent changes~\cite{situateRecSys, streamRecSysSurvey}. Other than the simple \textit{fine-tune} and \textit{full-retain} mentioned in Section~\ref{sec:intro}, researchers have proposed a few incremental learning strategies.  

\subsubsection{Experience Reply} It is a widely adopted retraining strategy~\cite{SPMF, SSRM, GAG, ADER}. The main idea is to maintain a reservoir of historical interactions to be retrained together with new interactions. SPMF~\cite{SPMF} keeps samples which cannot be predicted well, \ie hard samples, in reservoir. The key motivation is that hard samples require more training. 
SSRM~\cite{SSRM}, GAG~\cite{GAG} and Ader~\cite{ADER} adopt a similar approach for streaming session-based recommendation. Note that, although GAG uses graph convolution operation in training, the operation is  used to encode a session only, but not on the user-item interaction graph. Hence, it is a different topic from our work. Ader~\cite{ADER} stores exemplars for each item that is active in the current period. The number of exemplars of an item depends on the popularity of the item in the previous period. 
Apart from deep-learning based methods, the simple $k$-nearest-neighbors method can also achieve good results when handling streaming sessions~\cite{knn_vs_streaming}.

\subsubsection{Model-based Retraining} It is another widely adopted retraining strategy~\cite{SML, ASMG, DCMR}. SML~\cite{SML} and ASMG~\cite{ASMG} assume that model parameters learned using historical interactions well capture user preferences and item characteristics. The authors argue that it is not necessary to maintain a replay memory, to avoid catastrophic forgetting. Instead, they design transfer modules to combine the historical models and the new model learned from newly arrived interactions to predict future data. The transfer module is convolutional network in SML~\cite{SML}. In ASGM~\cite{ASMG}, the transfer module is a GRU network over a sequence of historical models in .

\subsubsection{Incremental Learning for GCN}  

The aforementioned retraining strategies are not designed for GCN-based models. In particular, the graph structure is not considered in retraining. 
GraphSAIL~\cite{graphsail} has three components to respectively preserve three pieces of information: local structure of a graph, global structure of a graph, and the self-information learned in previous period. 
Nevertheless, GraphSAIL utilizes the historical final representations as a whole without differentiating contributions from different neighbors in the current period. 
Solutions like CI~\cite{CI} and IGCN~\cite{IGCN} treat node representations from different layers differently. CI~\cite{CI} updates user and node representations  by aggregating each layer's old representations learned in the past and the new representation learned from the new interaction graph. 
Similar to CI, IGCN~\cite{IGCN} fuses old representation and new representation to capture both long-term and short-term information. CI uses convolutional neural network (CNN) in fusion, while IGCN uses temporal CNN network. 
Different from the aforementioned deep learning-based retraining framework, FIRE~\cite{FIRE} is a non-parametric incremental learning framework for graph-based recommendation. It uses a temporal graph filter to modify a user-item interaction matrix for graph convolutions technique. With this filter, historical interactions receiving lower weights during learning. 

Our proposed \modelname is more related to CI and IGCN. Both CI and IGCN fuse node representations learned in the previous period and that in the current period, layer by layer. That is, new and old representations learned from the same layer are fused. In some ways, neighbors in different hops are differentiated. However, we note that same layer representations learned from different graphs may reflect completely different neighbors, because one user-item interaction graph is constructed for each time period. For the same user, its neighbors will be very different across  different periods. Simply fusing representations learned from the same layer neglects the information differences across layers. Our \modelname is fundamentally different from CI and IGCN by design. We do not directly fuse node representations. Instead, we redesign the embedding layer of GCN to ensure that the initial features contain both old information and learnable new information. On top of that, we further introduce a disentanglement module to distinguish  old and new information, which is not done in CI and IGCN.

%% file: gcnmodel.tex
\section{Preliminary: GCN-based Model}
\label{sec:gcn}
Before we present the proposed \modelname, we briefly explain GCN-based recommendation model, which will be the base model for \modelname.  We focus on recommendations utilizing a \textit{user-item bipartite graph} - $\mathcal{G} = \left(\mathcal{V}, \mathcal{E}\right)$. That is, the nodes $\mathcal{V}$ in the graph are either \textit{user nodes} or \textit{item nodes}. The edge set $\mathcal{E}$ indicates the corresponding interactions between users and items. If a user interacts with an item, then there is an edge between them in the graph. Note that, there are no edges directly between users, or between items.

A GCN-based recommender model generally consists of (i) an embedding layer, (ii) $L$ neighbor aggregation \& information update layers,  and (iii) a final representation layer~\cite{GNNRecSys, GNNChallengesRecSys}.

\paratitle{Embedding layer.} Similar to other collaborative filtering based recommendation models~\cite{NeuMF, disentangleLongShort}, a graph-based recommender starts with an embedding layer. By user ID and item ID, the corresponding user embedding $e_u \in \mathbb{R}^d$ and item embedding $e_v \in \mathbb{R}^d$ can be obtained as the initial features of users and items. Here, $d$ is the dimension of an embedding vector. Both user and item embeddings are trainable, and are updated during model training.

\paratitle{Neighbor aggregation \& information update.} Given a graph and the features of user/item nodes, a GCN-based recommender allows a node to receive information from its neighbors. For example, message from neighbors $N(u)$ of a user node $u$ can be as follows.
\begin{equation}
    h_{N(u)} = AGGREGATION\left({h_v, \forall v\in N(u)}\right)
    \label{eq:neighbors_aggregation}
\end{equation}

To further utilize the graph structure, neighbor aggregation can be extended to enable high-order message propagation. That is, $L$ layers of neighbor aggregation can be stacked to allow for recursive drawing of information from $L$-hop neighbors. At the $\ell^{th}$ step, message received from user $u$'s neighbors is:
\begin{equation}
    h^{0}_v = e_v
\end{equation}
\begin{equation}
    h^{(\ell-1)}_{N(u)} = AGGREGATION\left({h^{(\ell-1)}_v, \forall v\in N(u)}\right)
    \label{eq:neighborsLayer_aggregation}
\end{equation}
Based on the aggregated message from neighbors, the user node representation at each propagation layer is then refined:
\begin{equation}
    h^\ell_u = UPDATE\left(\left\{h^{\ell-1}_u, h^{\ell-1}_{N(u)}\right\}\right)
\end{equation}

In the above description, we use user node as an example. The same aggregation and update processes apply to item nodes. Through the user-item bipartite graph, GCN-based recommender is able to model not only user-item relations, but also indirect user-user and item-item relations via the multi-hop aggregation.

\paratitle{Final representation layer.} By stacking $L$ message propagation layers, we obtain $L$ node representations, one at each layer. A recommender may decide how to derive a final representation for users/items. For instance, PinSage~\cite{pinsage} and STAR-GCN~\cite{star-gcn} choose the last-layer representation as the final representation $h^{final}_u$. Other models~\cite{lightgcn, ngcf, DGCF} aggregate all $L$ node representations as well as the initial embedding via techniques such as concatenation and mean pooling. 
\begin{equation}
    h^{final}_u = AGGREGATION\left( h^\ell_u, \forall \ell \in \left[0, \ldots, L\right] \right)
\end{equation}

The predicted preference between user $u$ and item $v$ is computed with the two final representations $h^{final}_u$ and $h^{final}_v$.

%% file: dilModel.tex
\section{The Proposed Model}

\begin{figure*}[t]
    \centering
    \includegraphics[trim = {4cm 4.5cm 4cm 1cm}, clip, width = 0.95\textwidth]{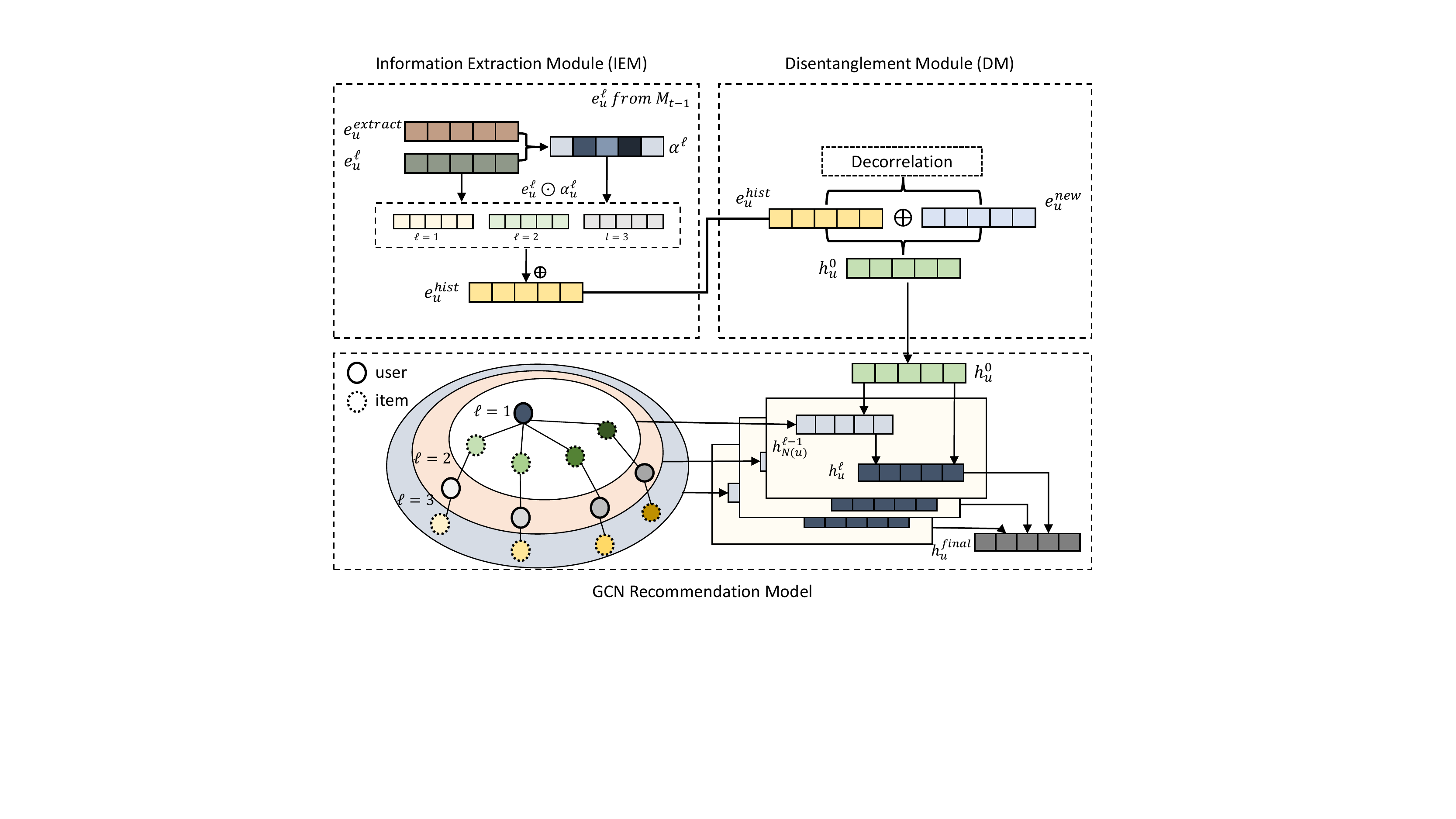}
    \caption{Model architecture of \modelname using user node as an example. \modelname consists of two components: Information Extraction Module and Disentanglement Module. The representations learned from \modelname can be used as input for any GCN-based recommendation model.}
    \label{fig:proposed_model}
\end{figure*}

We follow the model retraining process illustrated in Figure~\ref{fig:retraining_process}. Recall that, at the end of time $t_n$, model retraining has three pieces of information: (i) $R_n$ the newly observed user-item interactions within the time window $t_n$, (ii) $R_{<n}$ the historical interactions happened before $t_n$, and  (iii) $M_{n -1}$ the latest model deployed to service. With respect to the retrained model $M_n$, $M_{n-1}$ is also referred to as ``the previous model'' in our following discussion. 

As the newly observed interactions best reflect the current trends and are more relevant to near future recommendations, we build a new user-item bipartite graph by using $R_n$. To incorporate long-term preferences, we extract historical information from $M_{n-1}$, and then blend such information into the new model $M_n$. To avoid confusion, we present our proposed \modelname under the assumption that both users and items had interactions during $t_{n-1}$, which is applicable to majority users and items. At the end of this section, we detail how the other types of users/items are handled as special cases in our model. We will also discuss how the model deals with unseen users/items after being deployed to service in time $t_{n+1}$.

\paratitle{Architecture Overview.} In a user-item bipartite graph, the same retraining process applies to both user nodes and item nodes. Hence, we will only present the operations applied on user nodes to avoid duplication. Since there are only user nodes in our discussion, to ease presentation,  we simply use $e$ to denote user embeddings. When there is a need to distinguish the two, we use $e_u$ and $e_i$ to denote user and item embeddings, respectively. 

The overall model architecture of the proposed Disentangled Incremental Learning (\modelname) framework is shown in Figure~\ref{fig:proposed_model}.  To start, the Information Extraction Module (IEM)  extracts historical information $e^{hist}$ from representations learned in $M_{t-1}$. It represents the long-term user preference. Following that, the Disentanglement Module (DM) fuses $e^{hist}$ with $e^{new}$ to form the initial embedding $h^0$ of the user node in the newly built bipartite graph. The learnable $e^{new}$ denotes the dynamic user preference to be learned from the new user-item interactions in $R_n$, through model training. It can be initialized randomly or with reference to previous models. In our implementation, we initialize $e^{new}$ with the final representation $h^{final}$ learned in $M_{t-1}$. Here, we consider $h^{final}$ well reflects the recent recommendation context, which will be very relevant to the current time period. Nevertheless, it does not represent the long-term information, hence $e^{hist}$ remains essential in our model design. To ensure that the \modelname differentiates historical and new information, DM decorrelates $e^{hist}$ and $e^{new}$, shown in Figure~\ref{fig:proposed_model}. Then $h^0$ is used as the initial feature of user $u$ and fed to the base GCN recommendation model for training. In this manner, historical information is memorized and also differentiated from new information. Claimed as one of our contributions, we are the first to redesign the embedding layer in GCN-based recommender for model retraining. 

\subsection{Information Extraction Module (IEM)}
\label{ssec:iem}

As aforementioned, we consider the final representations learned in $M_{t-1}$ represent the recent recommendation context, but do not well represent the long-term preference of users and the persistent characteristics of items. Consider an example that a user switches to tennis from badminton. He or she remains  a sports lover in general. Although some recent interactions are more on badminton, but if we look at all his/her interactions and the related neighbors at multi-hops in the user-item graph, then the characteristics of sports lover are well demonstrated. The same applies to someone who upgrades phone to the latest version of the same brand. The past interactions with various phone accessories may well reflect the users' loyalty to this brand. Hence, we consider the user long-term preference and item persistent characteristics are not simple representations in the previous model. Rather, only partial information in the representations indicates long-term preference and persistent characteristics along time. Moreover, the long-term information is spread at different layers, in the form of aggregated information from multi-hop neighbors.

Based on the above understanding, we extract historical information from the representations at multiple layers from $M_{n-1}$. Note that $M_{n-1}$ is not limited to only the set of interactions $R_{n-1}$. It also retains historical information extracted from its previous model which in turn from $M_{init}$ by back-tracking. Our design aims to capture the slowly updated historical information across models along time.  

To determine what information to extract and how much information to extract from each layer, we design two parameters. We name the first parameter an \textit{extractor embedding} which determines the parts of information to be extracted from each node. The second parameter, \textit{weight vector}, then decides the amount of information to be extracted from each layer. In this work, we propose two implementation designs of the extractor embedding and weight vector. 

\paratitle{Design 1.} We initialize a \textit{learnable extractor embedding} $e^{extract}$ for each node to determine which historical features to extract. Then for each layer, we compute a \textit{weight paramter} to determine the weights of the features at that layer.  Let $e^\ell$ be the output (or aggregated information) at the $\ell$-th layer in $M_{n-1}$. Recall that  $e^\ell$ reflects the information gathered from the $\ell$-hop neighbors in time $n-1$, $\ell \in [0,1,\ldots,L]$. The weight for layer $\ell$ is computed as: 
\begin{equation}
    \alpha^\ell = \sigma\left(e^{extract} \odot e^\ell + pos^\ell\right);\quad \ell \in [0,1,\ldots, L]
    \label{eq:alpha_weight}
\end{equation}
Here, $\sigma$ is a Sigmoid function. It controls the feature extracted from each dimension to be in the range of $0$ to $1$. $\odot$ denotes element-wise product. In addition, $pos^\ell$ is a positional embedding which further controls the amount of information extracted from the previous period's $\ell$-hop neighbors.
With the weight $\alpha^\ell$ computed for each layer, we aggregate the extracted information by an $AGGREGATION(\cdot)$ function, which can be sum, mean pooling, or concatenation.

\begin{equation}
    e^{hist} = AGGREGATION\left(\alpha^\ell \odot e^\ell, \forall \ell \in [0,1, \ldots L]\right)
\end{equation}

\paratitle{Design 2.} Similar to design 1, we use $e^{extract}$ to determine the long lasting attributes in $e^\ell$ of each node, $\ell \in \left[0, \ldots, L\right]$. Instead of using a weight vector to determine the amount of information to extract, we initialize a learnable matrix $W^\ell \in \mathbb{R}^{d \times d}$ for each layer. $W^\ell$ directly distills useful information from the $\ell$-hop neighbors.

\begin{equation}
    e^\ell_{hist} = W^\ell(e^{extract} \odot e^\ell)
\end{equation}
Then we aggregate the information extracted from each layer by aggregation function.

\begin{equation}
    e^{hist} = AGGREGATION(e^\ell_{hist}, \forall \in [0, \ldots, L])
\end{equation}

We note that there are other possible designs for information extraction. One could always replace the proposed designs with a new extraction operation that cater to a specific domain, \eg fashion recommendation and video recommendation. As IEM is only responsible to extract $e^{hist}$, its design does not affect the overall architecture of DIL for incremental learning.  

\subsection{Disentanglement Module (DM)}
\label{ssec:dm}

Now we have the historical information $e^{hist}$ extracted for each node by IEM. $e^{hist}$ is information extracted from historical models. To capture the new recommendation context with the set of interactions $R_n$, we initialize an embedding $e^{new}$ for each node. To blend old and new information, we sum the two embeddings and obtain:

\begin{equation}
    h^0 = e^{hist} + e^{new}
\end{equation}
Then, $h^0$ will be used as the initial feature for a base GCN-based recommendation model, \eg LightGCN and NGCF. 

To ensure model retraining is able to distinguish $e^{hist}$ and $e^{new}$,  we introduce a disentanglement module. The key idea of disentanglement is to ensure that the base recommendation model learns as much additional information as possible on top of the historical knowledge $e^{hist}$. The additional information is new information reflected only by the new interactions.

For disentanglement, we conduct \textbf{decorrelation of representations}. Inspiring by~\cite{DGCF}, we adopt the distance correlation $CORR$ as a loss term, calculated by:
\begin{equation}
    L_{CORR}\left(e^{hist}, e^{new}\right) = \frac{dCov\left(e^{hist}, e^{new}\right)}{\sqrt{dVar\big(e^{hist}\big)\cdot dVar\big( e^{new} \big)}}
\end{equation}

In the above equation, $dCov(\cdot)$ denotes the distance covariance between two vectors while $dVar(\cdot)$ denotes the distance variance of an vector. Note that, the lower the $L_{CORR}$ values, the less dependent the input vectors are, and $L_{CORR}\in [0,1]$.  Via decorrelation, we hope that information learned in $e^{hist}$ and $e^{new}$ could be independent to each other, but both are useful in the message propagation with the final goal of more accurate  recommendation.

Both $e^{hist}$ and $e^{new}$ make up the initial embedding of a node. Along the training process, new interactions are used as ground-truth data instances to supervise the learning of both $e^{hist}$ and $e^{new}$. However, as $e^{hist}$ denotes long-term information existing in all time periods, historical interactions are necessary to provide additional supervision signal to $e^{hist}$. Hence, we propose to supervise the learning of $e^{hist}$ by historical interactions, detailed next.  

\subsection{Model Optimization}
\label{ssec:optim}

We initialize node representation $h^0$ with a fusion of $e^{hist}$ and $e^{new}$. Hence, we can have a user representation to be $h^0_u$, and an item representation to be $h^0_i$.  Both $h^0_u$ and $h^0_i$ are used as input for the base GCN recommendation model. After $L$ layers' message propagation as discussed in Section~\ref{sec:gcn}, we obtain the final node representations $h^{final}_u$ and $h^{final}_i$, for user and item respectively. Their dot product is used to predict a user's preference on an item.
\begin{equation}
    \hat{y}_{ui} = h^{final}_u \boldsymbol{\cdot}  h^{final}_i 
\end{equation}

We adopt pairwise BPR ranking loss~\cite{BPR} as our objective function. The key idea is that we hope to optimize the model parameters such that the predicted preference of a user is higher on the item that the user clicks/rates/purchases than an item that has no interactions with the user. We argue that clicks/rates/purchases behaviours demonstrate a user's interests on an item. In the incremental learning scenario, we update model parameters with the newly observed interactions in time $t_{n}$. In particular, we optimize the BPR loss $L_{new}$:
\begin{equation}
    L_{new} = \sum_{(u,i,j) \in O}-\ln \sigma
    \left(\hat{y}_{ui} - \hat{y}_{uj}\right)
    \label{eq:new_sup_loss}
\end{equation}
Item $i$ is an item interacted with user $u$ in the current period $t_n$, while item $j$ is a negative item sampled from the items that do not have interactions with user $u$ till $t_n$.

As discussed earlier, we further supervise $e^{hist}_u$ to ensure it well reflects historical information. 
To this end, we use historical interactions sampled from $R_{<n}$ to supervise them. Hence, we revise the loss term in Equation~\ref{eq:new_sup_loss} to:

\begin{align}
    L =\sum_{(u,i,k, j) \in O}-\ln \sigma\left(\hat{y}_{ui} - \hat{y}_{uj}\right) - \ln \sigma\left(e^{hist}_u \cdot e^{hist}_k  -  e^{hist}_u \cdot e^{hist}_j \right) \label{eqn:revisedL}
\end{align}

In Equation~\ref{eqn:revisedL}, item $k$ is an item sampled from the items that interacted with user $u$ in previous periods. Item $j$ is the sampled negative sample. Note that, as there is no interaction with user $u$, $j$ can be used as negative sample for training on both new interactions and past interactions. The additional supervision further ensures that the disentangled representation isolates the semantics of new and historical information.

Together with the disentanglement term, our ultimate objective function is as follows, where $\lambda$ is a hyperparameter which controls the weight on the disentanglement term.
\begin{equation}
    L_{final} = L + \lambda L_{CORR}
    \label{eq:final_loss}
\end{equation}

\subsection{Inactive Users and Inference}

In the incremental learning scenario, users can be categorized into three types. ``Active users'' have interactions in both  previous period $t_{n-1}$ and the current period $t_n$. Also, we have ``inactive users'' who have no interactions in $t_{n-1}$ but were active some time before $t_{n-1}$, Lastly, we have ``new users'' who have their first interactions in $t_n$. 

During the retraining process, we can obtain active users' information from $M_{n-1}$. For new users, their features are randomly initialized and will be updated in the current period. When it comes to inactive users, they have no historical information in $M_{n-1}$. There are three ways to initialize $e^{hist}$ for them. The simplest way is to treat them as new users. The second way is to extract $e^{hist}$ from the model when the user was last active. This would require to keep all historical model parameters. The last approach is to keep the inactive users in model retraining from the very beginning \ie $t_0$. Although they have no edges in the graphs constructed in their inactive periods, their parameters are updated along the model retraining. In our implementation, we used the last approach.

We note that active user is the largest group of users in a time period $t_n$, because users tend to visit a platform regularly. Taking the datasets we used in the experiments as examples, the average percentage of active users in a time period is 81.1\%, 64.3\%, 69.7\% in Amazon books, Amazon electronic, and Yelp respectively.

After retraining, the new model $M_n$ will be deployed for service in time $t_{n+1}$. There could be ``unseen users'' who have their very first interactions in time $t_{n+1}$. The feature of unseen users are randomly initialized and recommendation model will use the randomly initialized vectors for recommendation. In reality, if the platform is able to obtain certain attributes of the users \eg through registration, their features can be initialized based on the known attributes. The same applies to items.

%% file: Experiment.tex
\section{Experiment Setup}
\label{sec:exp}

We conduct experiments on three popular datasets - Amazon books~\cite{amazonDataset}, Amazon electronic~\cite{amazonDataset} and Yelp\footnote{\url{https://www.yelp.com/dataset}}. As \modelname is designed for GCN-based recommendation model, we select two graph-based recommendation models as our base model: LightGCN~\cite{lightgcn} and NGCF~\cite{ngcf}. Through experiments, we aim to verify the effectiveness and robustness of the proposed incremental learning framework. 

\begin{table*}[t]
    \centering
    \caption{Statistics of the datasets used in experiments.}
    \begin{tabular}[\columnwidth]{l|c|r|r|r|r}
        \toprule
        Dataset & Time Period & \#Users & \#Items & \#Warm-up interactions & \#Retraining interactions \\ 
        \midrule
         Amazon books & 01 Jan 2017 - 31 Dec 2017 & $34,195$ & $25,798$ & $458,145$ & $483,293$\\
         Amazon electronic & 01 Jan 2013 - 31 Dec 2017& $91,592$ &$34,401$ &$448,394$  & $1,028,395$ \\
         Yelp & 01 Jan 2012 - 31 Dec 2017 & $42,124$ & $29,099$ & $453,470$ & $649,627$ \\
         \bottomrule
    \end{tabular}
    \label{tab:dataStatistics}
\end{table*}

\subsection{Incremental Learning Scenario}
We follow the common setting for incremental learning in recommender systems~\cite{graphsail, SML, ASMG}. As illustrated in Figure~\ref{fig:retraining_process}, for each dataset, we set up a warm-up period to get a basic understanding of the users and items. The base recommendation model is batch-trained with all interactions happened during the warm-up period. 
Then, for all three datasets, we divide their interactions after the warm-up period into $6$ periods. The first set $R_0$ is used for retraining at $t_0$, then the updated model predicts  $R_1$. The prediction accuracy on $R_1$ is used as the validation metrics for hyperparameter tuning. The remaining $4$ sets of interactions $R_2$ to $R_5$ are treated as test sets. For each test set, the first 10\% interactions are used for early stopping, and we calculate the recommendation accuracy on the remaining 90\% interactions. The average recommendation accuracy over the $4$ test sets is the accuracy for a retraining strategy.

\paratitle{Datasets.}  The statistics of the datasets are reported in Table~\ref{tab:dataStatistics}. All three datasets have been preprocessed with $10$-core filtering, \ie all users and items have at least $10$ interactions. 

For \textit{Amazon Books}, we used the interactions from 2017. The first $6$ months of 2017 are set as the warm-up period. For the remaining $6$ months, each month's interactions constitute one set of data for retraining. For \textit{Amazon Electronic}, we select interactions from 2013 to 2017 for experiments. Out of the 5-year period, the first two years are set as the warm-up period. Then, we partition the remaining 3-year interactions into $6$ sets for retraining, with half-year data instances forming one set.  We select \textit{Yelp} interactions from 2012 to 2017 for experiments. The warm-up period is set to be 01 Jan 2012 to 31 Dec 2014. The interactions from 2015 to 2017 are then split into $6$ sets for retraining. Each set comprises interactions that happened within half a year. Listed in Table~\ref{tab:dataStatistics}, the number of interactions in warm-up period is comparable for all datasets.

\paratitle{Evaluation Metrics.} We formulate our task as a top-$N$ recommendation task with implicit feedback. Then we evaluate our proposed framework and baselines by the accuracy of top-$20$ recommendations. That is, at time $t_n$, we recommend $20$ items to each user, out of 
\textit{all items} seen till $t_n$.\footnote{For newly observed items in time $t_n$, there is no historical information for them. We randomly initialize their embeddings.}  The purpose of all-item-ranking~\cite{candidateSelections} is to avoid potential bias in recommendation results~\cite{sampledMetrics}. 

As for the evaluation metric, we report Recall@20 and NDCG@20.\footnote{Similar observations hold with Recall and NDCG on top-5 and top-10 recommendations. Hence, we only report Recall@20 and NDCG@20 for clarity.} Recall represents the percentage of correctly recommended instances among the test instances. NDCG (Normalized Discounted Cumulative Gain) further considers ranking positions of the correctly recommended item. Note that Recall@20 and NDCG@20 appear to be low in terms of their absolute values for two reasons. One is that we conduct all-item-ranking, and it is difficult to recommend a correct item among a large set of candidates. The other is that, the train/test data split strictly follows timeline in our setting. The recommendation accuracy is affected by both unseen users and unseen items in a new time window.

\subsection{Baselines}
We compare \modelname with five baselines. Among them, Fine-tune and Full-retrain are the two extreme retraining strategies. SPMF is an experience replay strategy; Both GraphSAIL and CI are specifically designed for GCN-based recommenders. GraphSAIL is distillation-based incremental learning framework while CI is a model-based incremental learning framework. To this end, we cover different types of incremental learning techniques in our experiments.

\paratitle{Fine-tune} retrains the base recommendation model using only the interactions observed in the current period. When fine-tuning the base model using the most recent interactions, we initialize the base model with the model parameters learned in previous period as a memorization of old information.

\paratitle{Full-retrain} retrains the base recommendation model using all interactions observed till the time when retraining is conducted.
    
\paratitle{SPMF} is a non-graph based incremental learning strategy which replays memory during the retraining process. Interactions which do not receive good recommendation in the past are considered samples which require additional training, thus being kept in memory for replay. We tune the memory size from $\{20000, 40000, 60000\}$. 
    
\paratitle{GraphSAIL} is a graph-based retraining strategy. It is designed with three distillation losses to preserve local structure and global structure of a graph, and to preseve self-information in the past. We tune the three distillation coefficients from $\{0.000001, 0.001, 1\}$. Moreover, GraphSAIL formulates global graph structure by $k$-means clustering on node representations learned from GCN-based recommendation model. We tune the number of clusters $k$ from $\{5,10,20\}$.

\paratitle{CI} is also a graph-based retraining strategy. It fuses old node representations learned in the past with new representation learned from newly observed interactions. Moreover, it updates representation of $k$ nearest inactive nodes although there are no interactions observed for these nodes. This is achieved by linking them with active nodes. We tune the number of neighbors in $k$-nearest neighbor from $\{10,30,50\}$. We tune the coefficient on inactive node updates from $\{0.25,0.5,0.75,1\}$. 

For all the models, we further tune learning rate from search space $\{10^{-6}, 10^{-5}, 10^{-4}, 10^{-3},10^{-2}\}$ and embedding dimensions from $\{32,64,128\}$. For regularization coefficient, we search from $\{10^{-6}, 10^{-4}, 10^{-2}\}$.

\subsection{DIL Implementation Details}
From grid search over the candidates of hyperparameters, we find that \textbf{Design 1} in the IEM module demonstrates better recommendation performance on Amazon books and Yelp, and on Amazon electronic with LightGCN. As for Amazon electronic with NGCF, \textbf{Design 2} is a better option. In the remaining part of the study, for each dataset, we report experiment results from their optimal designs. We train DIL by Adam optimizer with initial learning rates of $0.0001$ on both Amazon books and Yelp, and $0.00001$ on Amazon electronic. The regularization coefficient used for Amazon books and Yelp is $0.0001$, while it is $0.000001$ for Amazon electronic. We set the dimension size to be $128$ on all three datasets. The weights on the disentanglement term, \ie $\lambda$ in Equation~\ref{eq:final_loss}, are different across datasets and models. It is tuned from $\{10^{-6}, 10^{-5},10^{-4},10^{-3},10^{-2}\}$.

For base models, we set the final representation to be an aggregation over all layers. The aggregation function used is mean pooling. Although the original aggregation function in NGCF is a concatenation of all node representations, the authors~\cite{ngcf} mentioned that aggregation function is replaceable and we find that mean pooling leads to better recommendation performance.

\section{Experiment Results}

We compare the recommendation accuracy of DIL with five baselines on three datasets, using two base models. Then, we test the effectiveness of the two main components in DIL.

\begin{table*}[t]
    \centering
    \caption{Recall@20 and NDCG@20 results on Amazon books, Amazon electronic and Yelp, upon instantiating different retraining strategies on LightGCN and NGCF. Best results are in boldface and second best underlined.}
    \begin{tabular}{c|c|cc|cc|cc}
         \toprule
          \multicolumn{2}{c|}{Dataset} & \multicolumn{2}{c|}{Amazon books} & \multicolumn{2}{c|}{Amazon electronic} & \multicolumn{2}{c}{Yelp}\\
         \midrule
         Base Model & Baseline & Recall@20 & NDCG@20 & Recall@20 & NDCG@20 & Recall@20 & NDCG@20 \\
         \midrule
         \multirow{6}{*}{LightGCN}& Fine-tune & 0.0368 & 0.0140 & 0.0256 & 0.0099  & 0.0591 & 0.0237\\
        & Full-retrain & 0.0332 & 0.0116 & 0.0210 & 0.0080 & 0.0560& 0.0217\\
         & SPMF & 0.0459 & 0.0169 & 0.0262 & \underline{0.0101} & \underline{0.0627} & \underline{0.0249}\\
        & GraphSAIL & \underline{0.0490} & \underline{0.0180} & \underline{0.0265} & \underline{0.0101} & 0.0598& 0.0240\\
         & CI &   0.0390 & 0.0159& 0.0192 & 0.0071 & 0.0521& 0.0205\\
         & \textbf{DIL} &\textbf{ 0.0690} & \textbf{0.0263} & \textbf{0.0271}& \textbf{0.0102} & \textbf{0.0649} & \textbf{0.0259}\\
         
         \midrule
         \multirow{6}{*}{NGCF}& Fine-tune & \underline{0.0618}& 0.0221& 0.0206 & 0.0078 & 0.0355 & 0.0136 \\
        & Full-retrain &0.0285 & 0.0106 & 0.0187 & 0.0071 & 0.0362 & 0.0137\\
         & SPMF & 0.0582 & \underline{0.0224} & \underline{0.0265} & \underline{0.0102} & \underline{0.0445} & 0.0170 \\
        & GraphSAIL & 0.0590 & 0.0204 & 0.0240 & 0.0094 &0.0425 & 0.0165\\
         & CI & 0.0571 & 0.0200 & 0.0236 & 0.0091 & 0.0442 & \underline{0.0176}\\
         & \textbf{DIL} & \textbf{0.0635}& \textbf{0.0233}& \textbf{0.0274}& \textbf{0.0105}& \textbf{0.0478} & \textbf{0.0186} \\
         \bottomrule
    \end{tabular}
    \label{tab:exp_results}
\end{table*}

\subsection{Effectiveness \& Robustness}

Reported in Table~\ref{tab:exp_results}, on all  datasets, \modelname outperforms all baselines, measured by both Recall and NDCG. The version with LightGCN as base model outperforms its counterpart with NGCF on Amazon books and Yelp, by a large margin. Nevertheless, \modelname with LightGCN reports slightly poorer results on Amazon electronic than its NGCF version. GraphSAIL and SPMF are among the best performing baselines. In particular, GraphSAIL with LightGCN achieves the second best results on Amazon books and Amazon electronic. 

\begin{figure}
    \centering
    \includegraphics[width = 0.8\columnwidth]{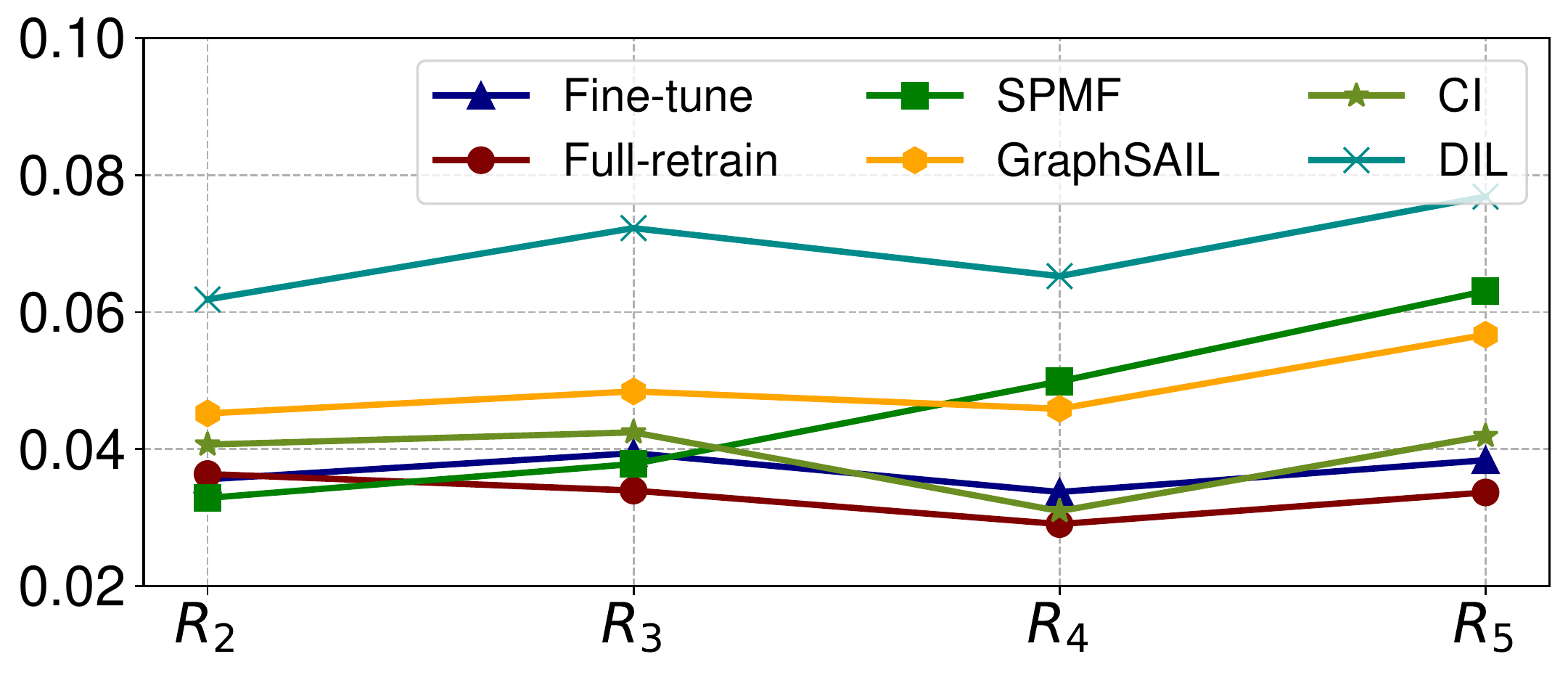}
    \caption{Recall@20 by retraining LightGCN on Amazon books, over the $4$ testing periods.}
    \label{fig:recall20_lightgcn_amazon_books_overPeriod}
\end{figure}

We further use LightGCN on Amazon books as an example and plot Figure~\ref{fig:recall20_lightgcn_amazon_books_overPeriod} to show the Recall@20 scores over the $4$ testing periods. An interesting observation is that performance ranking orders among the baselines vary across different periods. A similar observation is also made in~\cite{CVTT}, on other datasets. It calls for monitoring evaluation results over time to avoid potential bias in a particular time period. In this study, we average the evaluation scores over time and still observe better performance by \modelname. It clearly shows that \modelname is a robust and effective retraining framework.

Among all baselines, CI is considered the most similar to ours, because CI also fuses historical knowledge from $M_{n-1}$ with new information learned in the current period. The key difference lies in the ways of extracting and fusing the old and new information. That is, CIF conducts ``post-learning fusion''. It aggregates representations outputted by GCN in two different periods. Its extraction of historical information is straightforward. Unlike CI, \modelname conducts ``pre-learning fusion''. It fuses old and new information to be the input (initial feature of nodes) of GCN. By disentanglement, \modelname could well distinguish the historical and new information. Note that, during retraining, \modelname adjusts what historical information to be extracted by refining the information extraction module. 
Experiments show that DIL outperforms CI by a large margin.

Fine-tune outperforms Full-retrain under most settings except the version with NGCF on Yelp. This is a clear indication that the more recent interactions are more relevant to the current recommendation context. Nevertheless, Fine-tune does not show very competitive results in most cases (except Fine-tune on Amazon books with NGCF). It indicates that completely forgetting the historical information is not an ideal choice either. Both SPMF and DIL involve historical interactions sampled from $R_{<n}$ in training. In our experiments, we find that the optimal number of interactions needed by SPMF is not the largest option among the hyperparameter candidates. That also means a higher number of historical interactions may not necessarily lead to better recommendation.

Lastly, the absolute values on Amazon electronic are much lower than that on the other datasets. One reason is the sparsity of this dataset. Its user pool is two or three times larger than other datasets, while its number of interactions is only about 50\% more. Another reason is that the items in electronics may have a larger variety and shorter life span, compared to items in other categories \ie books and restaurants, making it a challenging dataset.

\subsection{Ablation Study}
\label{sec:ablation}

\begin{table*}[t]
    \centering
    \caption{Ablation study to verify the efficacy of designs in the Information Extraction Module.}
    \begin{tabular}{c|c|cc|cc|cc}
         \toprule
          \multicolumn{2}{c|}{Dataset} & \multicolumn{2}{c|}{Amazon books} & \multicolumn{2}{c|}{Amazon electronic} & \multicolumn{2}{c}{Yelp}\\
         \midrule
         Base Model & Baseline & Recall@20 & NDCG@20 & Recall@20 & NDCG@20 & Recall@20 & NDCG@20 \\
         \midrule
         \multirow{4}{*}{LightGCN}& DIL (mean) & \underline{0.0677} & \underline{0.0258} & \underline{0.0285} & \textbf{0.0111} & 0.0640 & 0.0256 \\
        & DIL (first) & 0.0613 & 0.0234 & 0.0269 & 0.0099 & \textbf{0.0658} & \textbf{0.0262}\\
         & DIL (last) & 0.0663 & 0.0197 & \textbf{0.0287} & 0.0085 & 0.0630 & 0.0187 \\
         & DIL & \textbf{0.0690} & \textbf{0.0263 }& 0.0271 & \underline{0.0102} & \underline{0.0649} & \underline{0.0259}\\         
         \midrule
         \multirow{4}{*}{NGCF}& DIL (mean) & \underline{0.0609} & 0.0229 & 0.0261 & \underline{0.0101} & 0.0435 & 0.0165 \\
        & DIL (first) & \textbf{0.0635}& \underline{0.0230} & 0.0183 & 0.0071 & \underline{0.0458} & \underline{0.0176} \\
         & DIL (last) & 0.0582 & 0.0221 & \textbf{0.0283} & 0.0084 & 0.0413 & 0.0160 \\
         & DIL & \textbf{0.0635} & \textbf{0.0233} & \underline{0.0274} & \textbf{0.0105} & \textbf{0.0478} & \textbf{0.0186} \\       
         \bottomrule
    \end{tabular}
    \label{tab:ablation_multiLayer}
\end{table*}

We further conduct experiments to verify the effectiveness of each component in DIL. Specifically, we explore different possible designs in the Information Extraction Module and the Disentanglement Module respectively. 

\paratitle{Information Extraction Module (IEM)} is designed to distinguish different neighbors and treat the $L$ layers' outputs differently. We conduct experiments to verify the needs of differentiating neighbors. In particular, we modify DIL to DIL (mean), DIL (first) and DIL (last). In DIL (mean),  we extract information from the mean node representations over all $L$ node representations and the initial embedding of a node. For DIL (first) and DIL (last), we extract historical information from the initial embedding and the last-layer representation respectively. 

Referring to the results in Table~\ref{tab:ablation_multiLayer},  different hops of neighbors provide different information, thus leading to inconsistent Recall@20 and NDCG@20. We also observe that ranking orders of DIL variants change across datasets and models. For example, when instantiating on LightGCN, mean node representation of GCN contributes more to recommendation performance on Amazon books, than using the initial embedding. When it comes to Yelp, the most important factor is the initial embedding. It indicates that neighbors should be given different weights depending on the domain of study. 

Our proposed DIL distinguishes neighbors from different hops in IEM. According to Table~\ref{tab:ablation_multiLayer}, our design is effective in most cases because DIL shows the best recommendation accuracy. As IEM is an independent module in DIL, it is not necessary to strictly follow the proposed designs. IEM can be revisited and redesigned to handle recommendations in different scenarios. Replacing IEM does not affect the overall architecture of DIL for retraining a GCN-based recommendation model.

\begin{table*}[t]
    \centering
    \caption{Ablation study to verify the efficacy of designs in the Disentanglement Module.}
    \begin{tabular}{c|c|cc|cc|cc}
         \toprule
          \multicolumn{2}{c|}{Dataset} & \multicolumn{2}{c|}{Amazon books} & \multicolumn{2}{c|}{Amazon electronic} & \multicolumn{2}{c}{Yelp}\\
         \midrule
         Base Model & Baseline & Recall@20 & NDCG@20 & Recall@20 & NDCG@20 & Recall@20 & NDCG@20 \\
         \midrule
         \multirow{3}{*}{LightGCN}& DIL-decorr & \underline{0.0617} & \underline{0.0231} & \underline{0.0255} & \underline{0.0095} & \underline{0.0643} & \underline{0.0258} \\
         & DIL-supervision & 0.0514 & 0.0192 & 0.0249 & 0.0092 &  0.0640 & 0.0257 \\
         & DIL-decorr-supervision & 0.0575 & 0.0208 & 0.0251 & 0.0093 & 0.0640 & 0.0251 \\
         
         & DIL & \textbf{0.0690} & \textbf{0.0263} & \textbf{0.0271} & \textbf{0.0102} & \textbf{0.0649} & \textbf{0.0259}\\         
         \midrule
         \multirow{3}{*}{NGCF}& DIL-decorr &  \underline{0.0620} & \underline{0.0229} & 0.0255&0.0098 & \underline{0.0468} & \underline{0.0181} \\
         & DIL-supervision & 0.0502 & 0.0182 & 0.0245 & 0.0095 &  0.0461& \underline{0.0181}\\
         
         & DIL-decorr-supervision & 0.0516 & 0.0188 & \underline{0.0262} & \underline{0.0101} & 0.0446 & 0.0172\\
         & DIL & \textbf{0.0635} & \textbf{0.0233} & \textbf{0.0274} & \textbf{0.0105}& \textbf{0.0478} & \textbf{0.0186} \\      
         \bottomrule
    \end{tabular}
    \label{tab:ablation_disentanglement}
\end{table*}

\paratitle{Disentanglement Module (DM)} decorrelates the long-term information and new information. The decorrelation is supervised by sampled historical interactions. We conduct experiments to explore changes in recommendation accuracy with removal of decorrelation and supervision from DIL. The results are reported in Table~\ref{tab:ablation_disentanglement}. 

The degradation of Recall@20 and NDCG@20 values in DIL-decorr and DIL-supervision compared to DIL, suggests that the decorrelation term and supervision are useful to better recommendations. On Amazon books, we observe that DIL-decorr-supervision performs better than DIL-supervison. It indicates that decorrelation may fail without supervision. One possible reason is that the model ends up with a trivial but not meaningful solution.

%% file: discussion.tex
\subsection{Discussion}
 We demonstrate the effectiveness of \modelname by comparing it with other incremental learning frameworks in experiments. We remark that, our experiments are designed to fairly evaluate all retraining strategies, with a simple retraining schedule at fixed time intervals. In reality,  retraining schedule is also a hyperparameter which largely affects recommendation accuracy. We further note that \modelname ignores $e^{hist}$ learned in other periods before time $t_{n-1}$, when making updates in the time window $t_n$, \eg ignoring $e^{hist}_{n-2}$ learned in time $t_{n-2}$. By definition, $e^{hist}_{n-2}$ contains information from $R_{n-3}$ as well as information from $R_{<n-3}$, while $e^{hist}_{n-1}$ contains information from $R_{n-2} \cup R_{<n-2}$. It can be inferred that $e^{hist}_{n-1}$ and $e^{hist}_{n-2}$ have intersections especially when there are long-lasting attributes which influence a user's behaviours across time periods. \modelname does not directly model the continuity of $e^{hist}$ for memorization of long-term preference, it maintains continuity by the supervision operation in the disentanglement module. It is not trivial to directly model continuity using $e^{hist}$ because $e^{hist}$ in different periods appear to be in different latent spaces. Moreover, continuity only exists in partial features of $e^{hist}$. We would leave the exploration as future work.